\documentstyle[preprint,aps,epsfig]{revtex}
\begin{document}
\title{Magnetic field-dependent interplay between incoherent and Fermi liquid transport mechanisms in low-dimensional
$\tau$ phase organic conductors}
%\title{Fermi Surface and electrical transport properties of a $\tau$ phase organic conductor}
\author{K. Storr$^1$, L. Balicas$^{1}$, J. S. Brooks$^1$, D. Graf$^1$, and G. C. Papavassiliou$^2$}
\address{$^1$National High Magnetic Field Laboratory, Florida State University, Tallahassee-FL 32306, USA}
\address{$^2$Theoretical and Physical Chemistry Institute, National Hellenic Research Foundation, Athens Greece 116/35}
\maketitle
\date{Received: \today }

\begin{abstract}
We present an electrical transport study of the 2-dimensional (2D)
organic conductor $\tau$-(P-({\em S,S})-DMEDT-TTF)$_2$(AuBr)$_2$
(AuBr$_2$)$_y$ (where $y \sim 0.75$) at low temperatures and high
magnetic fields. The inter-plane resistivity $\rho_{zz}$ increases
with decreasing temperature, with the exception of a slight
anomaly at 12 K. Under a magnetic field $B$, both $\rho_{zz}$ and
the in-plane resistivity plane $\rho_{xx}$ show a pronounced
negative and hysteretic magnetoresistance. In spite of a negative
residual resistivity ratio in zero field, Shubnikov de Haas (SdH)
oscillations are observed in some (high quality) samples above 15 T.
Furthermore, contrary to the single closed orbit Fermi surface
predicted from band structure calculations (where a single
star-shaped FS  sheet with an area of $\sim 12.5\%$ of $A_{FBZ}$
is expected), two fundamental frequencies $F_{l}$ and $F_{h}$ are
detected in the SdH signal. These orbits  correspond to $2.4 \%$
and $6.8\%$ of the area of the first Brillouin zone ($A_{BZ}$),
with effective masses $\mu_{l} = 4.0 \pm 0.5$ and $\mu_{h} = 7.3
\pm 0.1$ respectively. The angular dependence, in tilted magnetic
fields, of $F_{l}$ and $F_{h}$,  reveals a 2D character
of the FS, but no evidence for warping along the $k_z$ direction (
e.g., the absence of a beating effect in the SdH signal) is
observed. Angular dependent magnetoresistance (AMRO) further
suggests a FS which is strictly 2-D where the inter-plane hopping
$t_c$ is virtually absent or incoherent. The Hall constant
$R_{xy}$ is field independent, and the Hall mobility $\mu_{H}$
increases by a factor of $\sim 3$  under moderate magnetic fields.
Hence the field does not alter the carrier concentration, even in
the presence of a large negative magnetoresistance, but only
increases the lifetime  $\tau_{s}$. Our observations suggest a
unique physical situation where a stable 2D Fermi
liquid state in the molecular layers, are incoherently coupled along the
least conducting direction. The magnetic field not only reduces
the inelastic scattering between the 2D metallic layers, as seen
in the large negative magnetoresistance and SdH effect, but it
also reveals the incoherent nature of the interplane transport in
the AMRO spectrum. Finally, the observed Fermi surface is at odds
with band structure calculations. However, the very flat bands in
the electronic structure, combined with the variable charge
transfer, may be the origin of these effects. The apparent
ferro-magnetic character of the hysteresis in the
magnetoresistance, remains an unsolved problem.

\end{abstract}


%\twocolumn

\section{Introduction}

In the last two decades, the field of anisotropic low-dimensional organic
conductors has become synonymous with the observation of unusual and
exotic electronic properties. Examples range from the possibility of
unconventional,  anisotropic superconductivity \cite{1,2}, to the
observation of a variety of other ground states like charge-density waves (CDW) \cite{3}, spin-density waves (SDW)
\cite{4}, field-induced spin-density waves (FISDW) (associated to the
observation of quantum Hall effect \cite{5}), and the spin-Peierls state
(SP) \cite{6}. A considerable amount of effort has also been devoted to Fermiology \cite{7} and the
properties of the metallic states of these compounds. Non-Fermi liquid
like behavior has been reported  in photoemision spectra \cite{8}, there
are indications for spin-charge separation \cite{9} in some materials, and
unconventional electrical transport properties in the presence of magnet-field induced incoherent
hopping has been proposed \cite{10,11}.

More recently, new degrees of freedom are being added to these already
physically rich systems, by the incorporation of magnetic anions
into the structure of organic compounds.  Here, due to the physical
separation of the molecular orbital (cation) layers, and the inorganic
anion layers, there is a corresponding separation of the localized magnetic
anion moments (for example the $d$ electrons) and the itinerant low dimensional organic molecular
electrons gas ($\pi$ electrons). Typical examples are the series
$\lambda$-(BETS)$_2$Fe$_x$Ga$_{1-x}$Cl$_4$ compounds \cite{12} and TPP[Fe(Pc)(CN)$_2$]$_2$ \cite{13}. In the BETS
series, the progressive substitution of Ga by Fe suppresses the superconducting state and stabilizes an insulating anti-ferromagnetic (AF)
state \cite{12}. While in TPP[Fe(Pc)(CN)$_2$]$_2$, the ground state is also insulating and presents an anisotropic magnetic
susceptibility.

One of the main characteristics of magnetic organic systems as mentioned above, is the observation of a pronounced or giant
negative magnetoresistance under field. This effect has been explained in terms of field alignment of the local magnetic moments. On one hand, it is expected to destroy an eventual AF ground state, i. e., to close related gaps at the Fermi level (spin-flop transition) \cite{14},
and on the other, to decrease the spin scattering of itinerant electrons by these local moments. In any case,
the necessary ingredients for explaining the magnetic-field induced enhancement of the conductivity in these compounds, seems to be
the presence of localized magnetic moments, their interaction with itinerant electrons and the effects of the magnetic field
on this coupled system.

Nevertheless, there are other families of organic conductors, the
compounds of the $\tau$ crystallographic phase \cite{15}, whose
magnetoresistivity presents remarkable similarities to what is
observed, for example, in $\lambda$-(BETS)$_2$FeCl$_4$ and
TPP[Fe(Pc)(CN)$_2$]$_2$, although their structure is {\em not}
composed by any magnetic element, see Fig 1 a). Here we report on
the electrical transport properties of the $\tau$-(P-({\em
S,S})-DMEDT-TTF)$_2$(AuBr$_2$)$_1$(AuBr$_2$)$_y$ compound (where
$y \sim  0.75$ and P-({\em S,S})-DMEDT-TTF stands for
pyrazino-({\em S,S})-dimethyl-ethylenedithio-tetrathiafulvane), at
high magnetic fields $B$ and low temperatures $T$. In this
compound, an in-plane as well as inter-plane magnetoresistivity is
found to decrease by a factor $\geq 75 \%$ when a field $B \leq
10$ T is applied. A significant hysteresis is also observed
\cite{16}, which points towards the formation of field induced
domains and has been interpreted as an indication of the magnetic
nature of these compounds \cite{17}. However, magnetic
susceptibility measurements revealed an almost temperature
independent paramagnetic term. This term is comparable to those
measured in other 2D non-magnetic organic systems which
are characterized by strong electronic correlations \cite{18}.

The crystallographic structure of $\tau$-(P-({\em
S,S})-DMEDT-TTF)$_2$(AuBr$_2$)$_1$(AuBr$_2$)$_y$ is tetragonal
with unit cell dimensions {\bf a} = {\bf b} = 7.3546 $\AA$ and
{\bf c} = 67.977 $\AA$ \cite{18}. Inorganic anion layers alternate
with mixed organic-inorganic layers, which has both ordered and
disordered AuBr$_2$ anions, along with a disordered ethylene group
\cite{18}.  The ratio of donor molecules to acceptor anions is
$2:(1+y)$, where $y$ has been estimated to be $\sim 0.75$. The
value of $y$ determines the area of the Fermi surface, which
decreases with increasing $y$ \cite{19}.  Figure 1 b) shows the
calculated Fermi surface of $\tau$-(P-({\em
S,S})-DMEDT-TTF)$_2$(AuBr$_2$)$_1$(AuBr$_2$)$_y$, {$y \sim $0.75},
which was calculated using the extended H\"{u}ckel tight binding
method \cite{20}. The star shaped Fermi surface, results from the
four-fold symmetry of the molecules packing. While the {\bf a-b}
plane is metallic (conducting) the inter-plane electrical transport
displays an unusual non-metallic behavior in the whole temperature
range. This behavior contrasts with what is observed in most
quasi-two-dimensional (Q2D)organic compounds, where a $T^2$ behavior at
low $T$, is followed by a non-metallic behavior at higher
temperatures. A smooth crossover from coherent Fermi liquid
excitations at low temperatures, to incoherent excitations at high
temperatures, has been suggested to occur in these compounds
\cite{merino}.

In this paper, we report the first observation of Shubnikov de Haas (SdH) oscillations in a $\tau$ phase organic conductor;
the $\tau$-(P-({\em S,S})-DMEDT-TTF)$_2$(AuBr$_2$)$_1$(AuBr$_2$)$_y$ compound. Two fundamental frequencies $F_{l}$ and $F_{h}$
were detected in the fast Fourier transform of the SdH signal, corresponding respectively to $2.4 \%$ and $6.8\%$ of the area of
the first Brillouin zone ($A_{FBZ}$), which is at odds with band structure calculations. High effective masses,
$\mu_{l} = 4.0 \pm 0.5$ and $\mu_{h} = 7.3 \pm 0.1$ were obtained for $F_{l}$ and $F_{h}$, respectively.
The angular dependence of $F_{l}$ and $F_{h}$ reveals the 2D character of the FS
and the absence of frequency beatings, indicates that the FS is not warped along the $k_z$
direction. The angle dependent magnetoresistance (AMRO), suggests a strictly 2D FS, where the inter-plane
hopping $t_c$ is virtually absent or is incoherent. We find the Hall constant $R_{xy}$ to be field independent, and
the Hall mobility $\mu_{H}$ to increase by a factor of $\sim 3$, under moderate magnetic fields. This indicates that $B$ does not
introduce additional carriers into the system, instead, it decreases the carriers scattering rate $\tau_{s}^{-1}$.
As neither the inter-plane nor in-plane resistivity displays a $T^2$ dependence at zero field, we conclude, that
the magnetic field induces a crossover from a ``non-Fermi" liquid like behavior at moderate fields, towards a Fermi-liquid type behavior at higher fields, whose signature is the observation of quantum oscillations (QO's).

\section{Experimental Results}

Single crystals of $\tau$-(P-({\em S,S})-DMEDT-TTF)$_2$(AuBr)$_2$
(AuBr$_2$)$_y$ ($y \sim 0.75$), synthesized by electrochemical
methods \cite{papavassiliou1}, of which three different
morphologies, were used. Gold wires of 12.5 $\mu$m, were attached
with graphite paint, in a conventional 4 lead configuration for
inter-layer electrical transport measurements, see Fig 1 c), while
a 6 lead configuration was used for the Hall effect measurements.
Standard low frequency (~ 20 Hz) ac lock-in techniques, with
currents of order 10 $\mu$A were employed in the measurements.
Samples were mounted in a variety of fixed as well as rotating
sample holder probes, immersed in both $^3$He cryostats and
dilution refrigerators. Magnetic fields were provided by the
resistive magnets available at the National High Magnetic
Laboratory's DC field facility in Tallahassee Florida.

Figure 2 shows the typical temperature dependence at zero magnetic
field of the inter-plane resistivity $\rho _{zz}$, of a
$\tau$-(P-({\em S,S})-DMEDT-TTF)$_2$(AuBr)$_2$ (AuBr$_2$)$_y$ ($y
\sim 0.75$) single crystal, sample$\#$ 1. Although the in-plane
resistivity displays a metallic behavior \cite{papavassiliou2},
the inter-plane transport, as seen in the figure, is clearly
non-metallic and shows an abrupt change in slope at $T_{b} \simeq
12$ K. At this temperature, a metal-insulator transition has been
suggested to occur, although specific heat measurements did not
provide any evidence for a phase transition \cite{18} at $T_{b}$.
Below $T_{b}$, the in-plane resistivity presents a logarithmic
dependence on temperature \cite{papavassiliou2} at low $T$. This
temperature dependence has been interpreted as an indication of
either weak localization \cite{18,papavassiliou2} or Kondo effect
arising possibly from exchange interaction between localized
magnetic moments and itinerant conduction electrons
\cite{papavassiliou2}. In any case, and as clearly seen,
$\rho_{zz}$ does not display the typical $T^2$ dependence seen at
low $T$ in other Q2D organic compounds, which is the signature of
coherent electrical transport \cite{gorkov}.

Figure 3 shows the temperature dependence of $\rho _{zz}$ for $\theta \simeq 0^\circ $ ($\theta$ is the angle
between $B$ and the {\bf c} axis) and for several values of magnetic field $B$, as indicated in the figure. Several
striking features are observed:

i) $\rho _{zz}$ recovers a metallic character, i.e., $\rho_{zz}$ increases with $T$, below a magnetic field dependent temperature.

ii) For {\em all} values of magnetic field, $\rho _{zz}$ shows a crossover from positive to negative
magnetoresistance behavior at a crossover temperature $T_{a}(> T_{b}) \simeq 18$ K.

iii) The kink observed at $T_{a}$ is suppressed by the application
of a magnetic field.

The inset of Figure 3 shows the dependence of $\rho _{zz}$ on
temperature $T$ for two values of field, $B =$ 5 and 25 T
respectively, and for $\theta \simeq 0^\circ $. Arrows indicate
increasing and decreasing temperature sweeps. A marked $T$
dependent hysteresis is observed for $B=25$T which is similar to the behavior of magnetic-field induced domains. A small, but
non-negligible hysteresis is still observed at 5 tesla. The fact
that all the curves meet at $T_b > T_a$ suggests that $T_a$ does not
correspond to a thermodynamic phase transition. Instead, it could
indicate that charge transport in this system is described by
two distinct mechanisms with quite different temperature
dependencies. $T_a$ would correspond to the crossover temperature
between them. The mechanism that dominates at low temperatures
would be by some yet unknown reason and strongly magnetic field
dependent.

Figure 4 displays the magnetoresistance $R_{zz}$, from sample
$\#$1, as a function of magnetic field $B$ for $\theta \simeq
0^{\circ}$, and for four different temperatures: 1.45, 1.0, 0.7,
and 0.55 K respectively. All curves are vertically displaced for
clarity with arrows indicating field-up and field-down sweeps. As
previously reported \cite{17}, the resistance decreases by a
factor $\geq 65 \%$, followed again, by a significant temperature
dependent hysteresis. Furthermore, for $T \leq 1$ K and for fields
above $B \geq $ 17 tesla, Shubnikov-de Haas (SdH) oscillations are
observed. This is an indication of the high quality or long mean
free path at high fields of these $\tau$ phase metallic single crystals . 
The so-called resistivity ratio, $\Delta \rho = (\rho_{xx}
(300)-\rho_{xx}(4.2))/ \rho_{xx}(4.2)$, where $\rho_{xx}(300)$ is the
resistivity at $T=300$ K and $\rho_{xx}(4.2)$ is the resistivity at
4.2 K, is usually used as a criteria for judging the quality of an
organic metal. Typically, $\Delta \rho_{xx} \sim 100$ or greater is
found for most organic conductors that display SdH oscillations.
But this particular sample provides a value $\Delta \rho_{zz}
\simeq -0.9$ at $B=0$ tesla and $\Delta \rho_{zz} \simeq -0.8$ at
$B = 27$ tesla. This, at first glance, could be interpreted as a
clear indication of the ``low quality" or short mean free path of
our sample, {\em if}, SdH oscillations were not present.

Figure 5 shows the SdH signal as a function of inverse field
$B^{-1}$ for $\theta \simeq 0^{\circ}$ and for several values of
$T$, as indicated in the figure. The SdH signal is given by
$(\sigma - \sigma_{b})/ \sigma_{b}$, where $\sigma$ is the conductance or
the inverse of the actual resistivity of our sample (always valid if the
Hall component is small, which is the case for a metal), and $\sigma_{b}$
is the background conductance, obtained by inverting the background resistance.
$\sigma_{b}$ is obtained by fitting the actual sample resistance to a
third degree polynomial. The dotted line is a guide to the eye, showing the
envelope of the SdH signal. As clearly seen, it is necessary to postulate
{\em more than one} frequency for describing the SdH envelope through
the Lifshitz-Kosevich formalism. The inset of Fig. 5 shows the amplitude
of the fast Fourier transform (FFT) of the SdH signal corresponding to
$T = 0.55$ K as a function of frequency $F$. The FFT spectrum
displays two peaks at $F_{l} = $186 tesla and $F_{h} =$ 516 tesla,
respectively. The observation of two frequencies, i.e., two Fermi surface
extreme cross-sectional areas is surprising, since, according to band structure calculations
\cite{18}, the FS of this compound is composed only of a
closed star-shaped sheet (see Fig. 1b). Furthermore, from published crystallographic
data \cite{18}, the area of the first Brillouin (FBZ) zone is given by
$A_{FBZ} = 72.986$ nm$^{-2}$. Using the Onsager relation, $F = A(h/4 \pi^2 e)$,
where $F$ is the SdH frequency, $A$ the respective FS cross
sectional area, $e$ the electron charge and $h$ Planck's constant,
we obtained $2.4 \%$ and $6.8\%$ of the $A_{FBZ}$ for $F_{l}$ and $F_{h}$,
respectively. Nevertheless, the ratio of the area of the calculated closed
Fermi surface in Fig. 1 b) to $A_{FBZ}$ is estimated to be 1:8, corresponding
to a frequency of $F_{FS} = 955.8$ T. In other words, the  estimated $F_{FS}$
is considerably higher than either value determined in the present work.
The fraction of ``disordered" anions $y$, which determines the area of the FS
has been found to be time dependent in the $\tau$-(EDO-({\em S,S})-DMEDT-TTF)$_2$(I$_3$)$_{1+y}$ compound \cite{19}.
If $y$ in our sample were to differ from $\sim 0.75$, we
would expect the geometry of the FS to be quite different from what is shown in
Fig. 1a), perhaps this would explain the disagreement. The best refinement of the liquid crystal solution
of the compound $\tau$-(P-({\em S,S})-DMEDT-TTF)$_2$(AuBr)$_2$
(AuBr$_2$)$_1+y$ was found for y = 0.6 \cite{15}.
An alternative explanation could be that the ``kink" observed at $T_a$
is the onset of an eventual AF transition; as an AF transitions would
open partial gaps at the Fermi level and also affect the original geometry of the FS.
However, to date no indications of a phase transition at $T_a$ have not been found in specific heat or
magnetic susceptibility measurements \cite{16,18}.

Figure 6 shows the logarithm of the FFT amplitude of SdH signal
shown in Fig. 4 normalized with respect to $T$, as a function of
temperature for both frequencies $F_{l}$ (solid squares) and
$F_{h}$ (opened circles). The solid line is a fit to the
expression $X/ \sinh X$ where $X = \alpha \mu _{c} T/B$, $\alpha =
14.69$ T/K and $\mu_{c}$ is the effective cyclotron mass in
relative units of the free electron mass $m_e$. The slope yields the effective
cyclotron masses, $\mu_{l} = 4.0 \pm 0.5$ and $\mu_{h} = 7.3 \pm
0.1$ for $F_{l}$ and $F_{h}$, respectively. These are relatively high effective
masses for an organic metal, may not be surprising, since the
curvature of the proposed star-shaped Fermi Surface presents
singularities at its vertices. However, it could also suggest the
presence of localized magnetic moments: The exchange
interaction between carriers and localized moments are known to
modify dramatically the transport of carriers \cite{Hellman}, especially
near a metal-insulator transition. In general, complex
magnetoresistive behavior (combinations of positive and negative
magnetoresistivities as, for example, in manganites) has led to
theories for the formation of magnetic polarons \cite{kasuya}, i.e., ferromagnetic regions of local moments aligned with the spin
of the carrier, via the exchange interaction. Magnetic polarons
increase the carrier effective mass and hence tend to localize the
carrier, since it is dressed with a polarization ``cloud". At the
moment, there is no experimental evidence for this scenario.

Additional information can be obtained by the Lifshitz-Kosevich
formalism \cite{7} by plotting  the amplitude of the SdH
oscillations, normalized with respect to $B^{1/2}$. From this, we
obtain the Dingle damping factor: $R_{D} = \exp (- \alpha
\mu_{c}T_{D} / B)$ where $T_{D} = h /(4 \pi^{2}k_{B} \tau)$
($k_{B}$ is the Boltzmann  constant, $\mu_{c}$ is the carriers
effective mass in electronic mass units, and $\tau$ is the
relaxation time).  We calculated $T_D = 1.32 \pm 0.15$ K, which is
a small value typical of organic metals \cite{7}, and indicates
the high quality of this $\tau$ phase single crystal despite the
negative value of $\Delta \rho_{zz}$. Large effective masses make
it difficult to observe SdH oscillations for temperatures above 1
K, due to the Lifshitz-Kosevich damping factors.

Figure 7 displays the inter-plane magnetoresistance $R_{zz}$ as a function of magnetic field $B$ at $T \simeq 0.55$ K
and for several values of the angle $\theta$ (between $B$ and the inter-plane {\bf c} axis) as indicated in the figure.
Dotted arrows indicate field up and down sweeps. As indicated, the negative magnetoresistance, as well as the hysteresis
between field up and down sweeps, is markedly angle dependent. Apparently the higher the angle, the smaller the hysteresis.
Moreover, for $\theta > 50^{\circ}$ the quantum oscillations become virtually undetectable.

In figure 8 the angular dependence of both $F_{l}$ and $F_{h}$ is
plotted. Solid lines are fits to the expression $F(\theta)=
F(\theta = 0^{\circ})/ \cos {\theta}$. The fit is excellent and
provides values of $184 \pm 3$ and $520 \pm 6$ tesla for $F_{l}$
and $F_{h}$, respectively. It also clearly indicates that the FS
of the $\tau$-(P-({\em S,S})-DMEDT-TTF)$_2$(AuBr)$_2$
(AuBr$_2$)$_y$ ($y \sim 0.75$) compound is 2-dimensional as
expected for an anisotropic layered compound.

Figure 9 displays $R_{zz}$ as a function of $\theta$ for two
values of the in-plane angle $\phi = 0^{\circ}$ (solid line) and
$\phi = 45^{\circ}$ (dotted line) at $T = 4.2$ K and $B = 14$
tesla. As indicated by both schemes in the figure, $\phi$ is
defined with respect to one of the sample's edge, consequently,
$\phi = 45^{\circ}$ corresponds to a rotation along one of the
diagonals of the square-shaped sample. There is no sign of the
Yamaji effect, instead, at $\theta = 0^{\circ}$ there is an
incoherent peak. Furthermore, at $\theta  = \pm 90^{\circ}$, the
$\phi$ dependence indicates a 4-fold symmetry which at high fields
agrees with what is expected from Fig. 1b.  There, the resistance
peaks have a period of $\phi = 90^{\circ}$ for the in-plane MR.
The observation of a central peak in $R_{zz}$, i.e., for $B
\parallel I \parallel {\bf c}$, is quite surprising, since
magnetoresistance is not expected under these conditions according
to classical transport theory. In fact, in most 2D
organic compounds, peaks are observed at $\theta = 90^{\circ}$ (as
in our data for $\phi = 45^{\circ}$) and also with a well defined
periodicity in $\tan \theta$ \cite{1,7}. This periodicity,
according to a semi-classical approach \cite{yamaji}, results from
the warping of FS along the $k_{z}$ reciprocal lattice direction.
More recently, in the frame of an incoherent interlayer transport
model, McKenzie and Moses \cite {rossandmoses} demonstrated that
the existence of a three-dimensional (3D) FS is {\em not} a necessary
ingredient to explain the above mentioned angle dependent
structures which are observed in the magnetoresistance of most
organic conductors. In any case, the beating between two close
frequencies corresponding to the two FS extremal cross sectional
areas is a clear signature of a 3-D closed orbit which is warped
along the $k_z$ direction. This beating is absent in our data. An
indication of a finite inter-layer transfer integral is the
observation of a sharp and pronounced peak at $\theta =
90^{\circ}$ \cite{ross,hanasaki2}. In our data, a broad peak
showing a maximum at $\theta = 90^{\circ}$ is observed for $\phi =
45^{\circ}$ but is absent for $\phi = 0^{\circ}$, which may
suggest a strictly 2-D FS for these $\tau$ phase organic
compounds. Furthermore, our preliminary angular studies indicate
that the four-fold symmetry observed for $\phi = 45^{\circ}$,
decreases to a two-fold one at lower fields. A detailed angular
study will be the subject of a future report \cite{brooks}.

The in-plane resistivity, $\rho _{xx}$, as a function of $B$ from
sample $\# 2$ and for four different temperatures is presented in
figure 10 a). The respective temperatures are indicated in the
figure. In order to measure the Hall effect, a conventional 6 lead
configuration was used. The general behavior of $\rho_{xx}$ is
essentially similar to what is observed in $\rho_{zz}$ under
field: A remarkable resistivity drop, which in this case, is
observed for $B \leq 2$ T instead of $B \leq 6$ T as for
$\rho_{zz}$. Nevertheless, for this particular sample no quantum
oscillations are observed, which is surprising since, this single
crystal originally is from the same electrocrystallization cell as
sample $\# 1$. This clearly indicates that single crystals of
different qualities and/or physical properties are produced during
the synthesis process \cite{papavassiliou3} and may also suggest
that the content of acceptor anions, $1 + y$, may vary from sample
to sample in a single batch. Also, the morphology of this sample
differs from what is shown in Fig. 1 c), in this case we have
chosen a thin rectangular platelet. Figure 10 b) shows the Hall
resistance $R_{H}$ as a function of $B$ and for the four values of
$T$ in figure 10 a). $R_{H}$ is obtained by anti-symetrization of
the Hall voltage $V_{H}$: $R_{H} \equiv [V_{H}(+B) -
V_{H}(-B)]/2I_{x}$ where $I_{x} = 50 \mu$A is the injected
current. As seen, $R_{H}$ is linear in field, as expected for a
metal characterized by only one type of carrier, whose sign
indicates that electrons are the charge carriers, in agreement
with previous results \cite{Fortune}. Moreover, $R_{H}$ is
temperature independent below 4.2 K (solid line in this figure is
a guide to the eye). The Hall constant $R_{xy} = E_{y}/j_{x}
\equiv (R_{H} \cdot t/B)$, where $E_{y}$ is the transverse
electric field and $j_{x}$ is the in-plane density of current, is
shown as a function of $B$ in figure 11 a), from the traces in
Fig. 10 b). Except at low fields, where the Hall signal is too
small for an accurate determination, $R_{xy}$ is essentially
constant in magnetic field for $B$ up to 30 T. In other words,
there is no clear evidence which could indicate that $B$
introduces carriers into the system, hence decreasing its
resistivity. An estimation of the density of carriers in our
system is provided by the standard expression for the Hall
coefficient in a isotropic system: $n = (R_{xy} \cdot e)^{-1}$
where $e$ is the electron charge. $n$ is presented in Fig. 11 b)
and is basically constant for $B>4$ T saturating to a value $n
\simeq 3.75 \times 10^{26}$ m$^{-3}$. By multiplying $n$ by the
unit cell volume $\Omega = 3676.9$ $\AA ^{3}$ we obtain a value of
$\simeq 1.4$ carriers per unit cell. This value is remarkably
close to the number of acceptor anions $1 + (y \simeq 0.75)=1.75$
$\Omega^{-1}$, considering the usual uncertainty associated with
the sample and contacts geometrical factors as well as the
inadequacy of the above expression for describing a temperature
dependent Hall effect in an anisotropic 2-D system. Consequently,
and at least at low temperatures, the number of carriers seems to
be given by the number of acceptor anions in this $\tau$ phase
system. Finally, as $R_{xy} \ll \rho _{xx}$, the Hall mobility,
which is proportional to $\tau_{s}$, the inverse of the scattering
rate, is approximately given by $\mu_{H} \simeq R_{xy}/ \rho_{xx}$
and is plotted in figure 11 c). As seen, $\mu_{H}$ is rather
small, on the order of $10^{-2}$, and slightly decreases with $B$
indicating that $\tau _{s}$ increases at higher fields. In the
important low field region, where the resistance decreases
considerably, it is not possible to directly extract the real
behavior of $\mu_{H}$ due to the uncertainties in $R_{xy}$, as
mentioned above. Nevertheless, as $R_{H}$ is remarkably linear in
field, we expect $R_{xy}$ to be essentially constant in the whole
field range. As the resistivity decreases by a factor of ~ 3,
$\mu_{H}$ necessarily {\em increases} by the same factor. At the
moment, it is not clear which mechanism is responsible for this
magnetic-field induced reduction of $\tau_{s}^{-1}$.

\section{Discussion}

It would be possible to explain most of our observations, if we
assume the presence of localized spins that undergo an AF
transition at $T_b$, for example, ferromagnetic layers interacting
antiferromagnetically. The negative magnetoresistance, could be
ascribed to a decrement in the itinerant carriers scattering rate
$\tau_s$ (in agreement with our Hall effect measurements) due to
the alignment of the localized spins, or simply to the field
suppression of an insulating antiferromagnetic state as is seen in
several magnetic organic compounds. The presence of ferromagnetic
domains and/or hysteretic behavior would naturally be explained by
this scenario. In addition, and as mentioned before, the
discrepancy between the calculated Fermi surface and the SdH
frequencies found by us, could be simply explained in terms of a
FS reconstruction at $T_b$. On the other hand, the heavy carrier
effective masses would be the result of an indirect exchange type
of interaction, through ferromagnetic polarons. Nevertheless, and
as previously mentioned, this scenario is unrealistic, since none
of the constitutive elements in $\tau$-(P-({\em
S,S})-DMEDT-TTF)$_2$(AuBr)$_2$ (AuBr$_2$)$_y$ ($y \sim 0.75$) is
magnetic. At this point, to explain the negative magnetoresistance
seen here, it is interesting to mention the model for repulsively
interacting electrons on a lattice whose band dispersion contains
a flat portion, recently proposed by Arita {\em et al.}
\cite{arita}. According to these authors, when the Fermi level
lies in the flat part, electronic correlations cause ferromagnetic
spin fluctuations and consequently an enhanced spin
susceptibility. So far, the only experimental report \cite{18}
containing magnetic susceptibility measurements do not seem to
sustain this scenario.

The observation of an important hysteresis in the field dependence
of the resistivity and which points towards metastability in the
system, might provide some insight about the nature of the ground
state. Metastabilitiy is well known in density-wave (DW) type
systems and is associated with the DW pinning to, for example,
defects or the lattice. DW ground states are associated with the
geometry of the FS \cite{1} through the so-called FS nesting
property, which is a particularly useful concept for
quasi-one-dimensional systems with opened FS's. However, in our case,
we can not identify a ``good" nesting wave-vector for the
star-shaped FS resulting from band structure calculations (see
Fig. 1 b)). Also, negative magnetoresistance is compatible with a
charge-density wave (CDW) ground state which is destabilized by
the Pauli effect under magnetic field \cite{Pauli,jeremy}. Assuming
that $T_b \sim 13$ K is the CDW transition temperature and using a
simple BCS relation, we can estimate the critical field necessary
to suppress a uniform CDW: $B_{c}=1.765 k_{B}/ \mu_{\rm B}  T_{c}$
\cite{jeremy}, where $k_{B}$ is the Boltzmann constant, $\mu_{\rm
B}$ is the Bohr magneton, and $T_{c}$ is the transition
temperature to the CDW state. We obtain a critical field $B_{c}
\sim 43$ T for suppressing an eventual CDW ground state, which is
one order of magnitude higher than the magnetic fields at which a
pronounced resistivity decrement is observed by us. In our
opinion, a DW type ground state does not seem to be compatible
with our observations.

Finally, and as already indicated \cite{18,papavassiliou2} the resistivity at zero field, presents a $\ln T$ dependence, which
suggests a weak localization type of regime. Nevertheless, it is difficult to reconcile weak localization resulting from weak
disorder, with the angular dependent hysteresis seen here. Hysteresis would be compatible with highly disordered systems, where
the magnetic field is expected to affect, for example, the spin-orbit coupling, the configuration of localized charges/spins and/or
the percolation path for electrical transport in the system.

The lack of a $T^2$ dependence in the resistivity and/or frequency
beats in the SdH signal as well as of a peak in the
angular-dependent magnetoresistance for in-plane magnetic fields,
clearly suggests that the inter-plane transport is incoherent in
this $\tau$ phase organic conductor. Consequently, this compound
may be classified as a highly correlated low dimensional
electronic system, as is the case for most transition metal
oxides, like, for example, the cuprates, and for which an
appropriate physical description is yet unavailable. What is
really remarkable in the present case, is the observation of
Fermi-liquid type of behavior at high fields. Consider, for
example, the model developed in Refs. \cite{11,clark}, for
explaining, for example, the angular dependence of the
magnetoresistance in (TMTSF)$_2$PF$_6$. According to these
authors, fields applied {\em exactly} along the inter-plane axis
of an anisotropic, for instance, layered material do {\em not}
affect the inter-plane motion of the charge carriers, i. e., the
coherence of the inter-plane transport. Consequently, if the
system displays FL behavior at zero field, it should keep its
physical properties at this particular orientation under field.
But, for any other orientations, the magnetic field is expected to
interfere the interchain coherence by ``adding an effective
inelasticity to the interchain hopping \cite{11}", and
consequently, a non-FL type of behavior should emerge. Now,
consider what is observed in Fig. 7 where SdH oscillations are
still observed at $T \simeq 0.55$ K for angles as a pronounced as
$\theta = 37.3^{\circ}$ and for fields as high as 30 tesla. In
other words, even for an in-plane field component as high $\sim
18$ tesla this system, {\em recovers} its FL like character. A simple
way to conciliate our results with this model, is to assume that
inter-layer transport is incoherent at zero field and {\em remains} incoherent
in the whole field range. This could explain the absence of beatings
in the SdH signal, the absence of a sharp peak for $\theta = 90^{\circ}$ in the AMRO and
the negative resistivity ratio $\Delta \rho_{zz}$, or the unusual temperature
dependence of $\rho_{zz}$. Consequently, to assume that the transport is {\em coherent}
within the conducting planes is the only possible way to explain the observation
of quantum oscillations at high fields. This would imply an unusual physical situation:
2-D Fermi liquid type of layers coupled incoherently between them. However,
and as already mentioned, $\rho_{xx}$ does not diplay a $T^2$ dependence at low $T$ either.
Consequently, it would still be necessary to find an explanation for the observed
crossover towards FL behavior, within the conducting planes, induced by high magnetic fields.
The field induced reduction of $\tau_{s}^{-1}$ is a physical evidence for this crossover.

\section{Summary}

In summary we presented an electrical transport study in the
2-dimensional organic conductor $\tau$-(P-({\em
S,S})-DMEDT-TTF)$_2$(AuBr)$_2$ (AuBr$_2$)$_y$ (where $y \sim
0.75$) at low temperatures and high magnetic fields $B$. Both the
in-plane and the inter-plane resistivities show a pronounced
negative and hysteretic magnetoresistance, which in some samples
are followed by the observation of Shubnikov de Haas (SdH)
oscillations. Two fundamental frequencies $F_{l}$ and $F_{h}$ were
detected in the FFT spectrum of the SdH signal, corresponding
respectively to $2.4 \%$ and $6.8\%$ of the area of the first
Brillouin zone ($A_{FBZ}$), which disagrees with band structure
calculations. High effective masses $\mu_{l} = 4.0 \pm 0.5$ and
$\mu_{h} = 7.3 \pm 0.1$ were obtained for $F_{l}$ and $F_{h}$,
respectively. The angular dependence of $F_{l}$ and $F_{h}$
reveals the 2-dimensional character of the FS, while the absence
of frequency beatings indicates the absence of warping along the
$k_z$ direction. Furthermore, the angle dependent
magnetoresistance (AMRO) suggests a FS which is strictly 2-D, i.
e., the inter-plane hopping $t_c$ is incoherent. While the Hall
constant $R_{xy}$ is field independent, the Hall mobility
$\mu_{H}$ increases by a factor of $\sim 3$, under moderate
magnetic fields. This indicates that $B$ does not introduce
carriers into the system but, instead, decreases the carriers
scattering rate $\tau_{s}^{-1}$.

Although these observations might be explained in terms of the presence of local magnetic moments in the system, this possibility
is discarded by magnetic susceptibility measurements. These observations are also difficult to conciliate with a density-wave
type of ground state or with an explanation in terms of weak localization. Considering that there is no evidence for a thermodynamic
phase transition at $T_{b}$, we believe that we have found an example of a very anisotropic
Q2D system which shows a crossover from a non-Fermi liquid type of behavior to a Fermi liquid one,
induced by high magnetic fields. Considering the accumulation of evidence for incoherent charge transport between the layers,
 the only possible way to explain the observation of quantum oscillations at high fields, is to assume that
the transport is {\em coherent}, or FL like, within the conducting planes. Consequently, this $\tau$ phase
compound seems to be described by a unique scenario: Conducting 2-D layers displaying Fermi liquid type of behavior,
which are coupled incoherently.  The future confirmation of this scenario will certainly be relevant
for other layered materials like the transition metal oxides.

\section{Acknowledgments}

We are indebted to V. Dobrosavljevic and K. Murata for helpful discussions and S. MacCall for his help with the SQUID measurements.
We also acknowledge support from NSF-DMR 95-10427 and 99-71474 (JSB). One of us (LB) is grateful to the
NHMFL for sabbatical leave support. The NHMFL is supported through a cooperative agreement between
the State of Florida and the NSF through NSF-DMR-95-27035.

\begin{figure}[htbp]
%\begin{center} %\epsfig{file=taufig1.eps,width=75mm}
\caption{a)A sketch of the P-({\em S,S})-DMEDT-TTF molecule. b)
Calculated Fermi surface of $\tau$-(P-({\em
S,S})-DMEDT-TTF)$_2$(AuBr)$_2$ (AuBr$_2$)$_y$ and for $y \sim
0.75$ (solid line). The dashed line represents the calculated
Fermi surface for $y = 0$. c) Configuration of contacts for
inter-plane electrical transport
measurements, and morphology of the sample used.} %\end{center}
\end{figure}

\begin{figure}[htbp]
%\begin{center}
%\epsfig{file=taufig2.eps, width=8.5cm}
\caption{Dependence on temperature $T$ of the inter-plane
resistivity $\rho _{zz}$ of a $\tau$-(P-({\em
S,S})-DMEDT-TTF)$_2$(AuBr)$_2$ (AuBr$_2$)$_y$ ($y \sim 0.75$)
single crystal at zero magnetic field. The change in slope
observed at $T_{a}$ may indicate either a phase transition or a
crossover between two different electrical transport regimes.}
%\end{center}
\end{figure}

%\pagebreak
\begin{figure}[htbp]
%\begin{center} \epsfig{file=taufig3.eps, width=8.5cm}
\caption{Inter-plane resistivity $\rho_{zz}$ as a function of
temperature $T$ under several values of magnetic field applied $B$
along the inter-plane {\bf c} axis. For {\em all} values of $B$,
$\rho _{zz}$ shows a crossover from positive to negative
magnetoresistance behavior at a crossover temperature $T_{a}(>
T_{b}) \simeq 18$ K. Inset: $\rho_{zz}$ for both field up and down
sweeps (indicated in the figure by arrows) as a function of $T$
and for two values of field 5 and 25 T, respectively. A large
hysteresis is observed at 25T.} %\end{center}
\end{figure}

\begin{figure}[htbp]
%\begin{center} %\epsfig{file=taufig4.eps, width=7.5cm}
\caption{Inter-plane resistivity $\rho_{zz}$ (sample $\#$1) as a
function of $B$ for $\theta \simeq 0^{\circ}$ and for four
different temperatures: 1.45, 1.0, 0.7, 0.55 K, respectively.
Curves are vertically displaced for clarity. Arrows indicate
field-up and field-down sweeps} %\end{center}
\end{figure}

\begin{figure}[htbp]
%\begin{center} %\epsfig{file=taufig5.eps, width=8.5cm}
\caption{SdH signal as a function of inverse field $B^{-1}$ for
$\theta \simeq 0^{\circ}$ and for several values of $T$. Dotted
line is a guide to the eye. Inset: The amplitude of the fast
Fourier transform (FFT) of the SdH signal corresponding to $T =
0.55$ K as a function of frequency $F$. Two peaks are observed at
$F_{l} = 186$ T and $F_{h} = 516$ T, respectively.} %\end{center}
\end{figure}

\begin{figure}[htbp]
%\begin{center} \epsfig{file=taufig6.eps, width=8.5cm}
\caption{The logarithm of the FFT amplitude of the SdH signal
previously shown in Fig. 5 normalized respect to $T$, as a
function of temperature for both frequencies $F_{l}$ (solid
squares) and $F_{h}$ (opened circles). Solid lines are fits to the
Lifshitz-Kosevich formalims that provides the effective masses
$\mu_{l}= 4.0 \pm 0.5$ for $F_l$ and $\mu_h = 7.3 \pm 0.1$ for
$F_h$, respectively. } %\end{center}
\end{figure}

\begin{figure}[htbp]
%\begin{center} %\epsfig{file=taufig7.eps, width=8.5cm}
\caption{Inter-plane magnetoresistance $R_{zz}$ as a function of
$B$ at $T \simeq 0.55$ K and for several values of the angle
$\theta$ between $B$ and the inter-plane {\bf c} axis. Dotted
arrows indicate field up and down sweeps.} %\end{center}
\end{figure}

\begin{figure}[htbp]
%\begin{center} %\epsfig{file=taufig8.eps, width=8.5cm}
\caption{Angular dependence of both $F_{l}$ and $F_{h}$. Solid
lines are fits to the expression $F(\theta)= F(\theta =
0^{\circ})/ \cos {\theta}$.} %\end{center}
\end{figure}

\begin{figure}[htbp]
%\begin{center} %\epsfig{file=taufig9.eps, width=8.5cm}
\caption{$R_{zz}$ as a function of $\theta$ for two values of the
in-plane angle $\phi = 0^{\circ}$ (solid line) and $\phi =
45^{\circ}$ (dotted line) at $T = 4.2$ K and $B = 14$ tesla.}
%\end{center}
\end{figure}

\begin{figure}[htbp]
%\begin{center} \epsfig{file=taufig10.eps, width=8.5cm}
\caption{a) The in-plane resistivity $\rho _{xx}$ as a function of
$B$ from sample $\# 2$ and for four different temperatures. b) The
Hall resistance $R_{H}$ as a function of $B$ for the same four
values of $T$ as in a). The solid line is a guide to the eye.}
%\end{center}
\end{figure}

\begin{figure}[htbp]
%\begin{center} \epsfig{file=taufig11.eps, width=7.5cm}
\caption{a) The Hall constant $R_{xy}$ (see text) as a function of
$B$ calculated from the traces shown in Fig. 10 b). b) The density
of carriers $n$ as a function of $B$ obtained from the traces in
Fig. a). c) The Hall mobility $\mu_{H} \simeq R_{xy}/ \rho_{xx}$
as a function of field $B$.} %\end{center}
\end{figure}

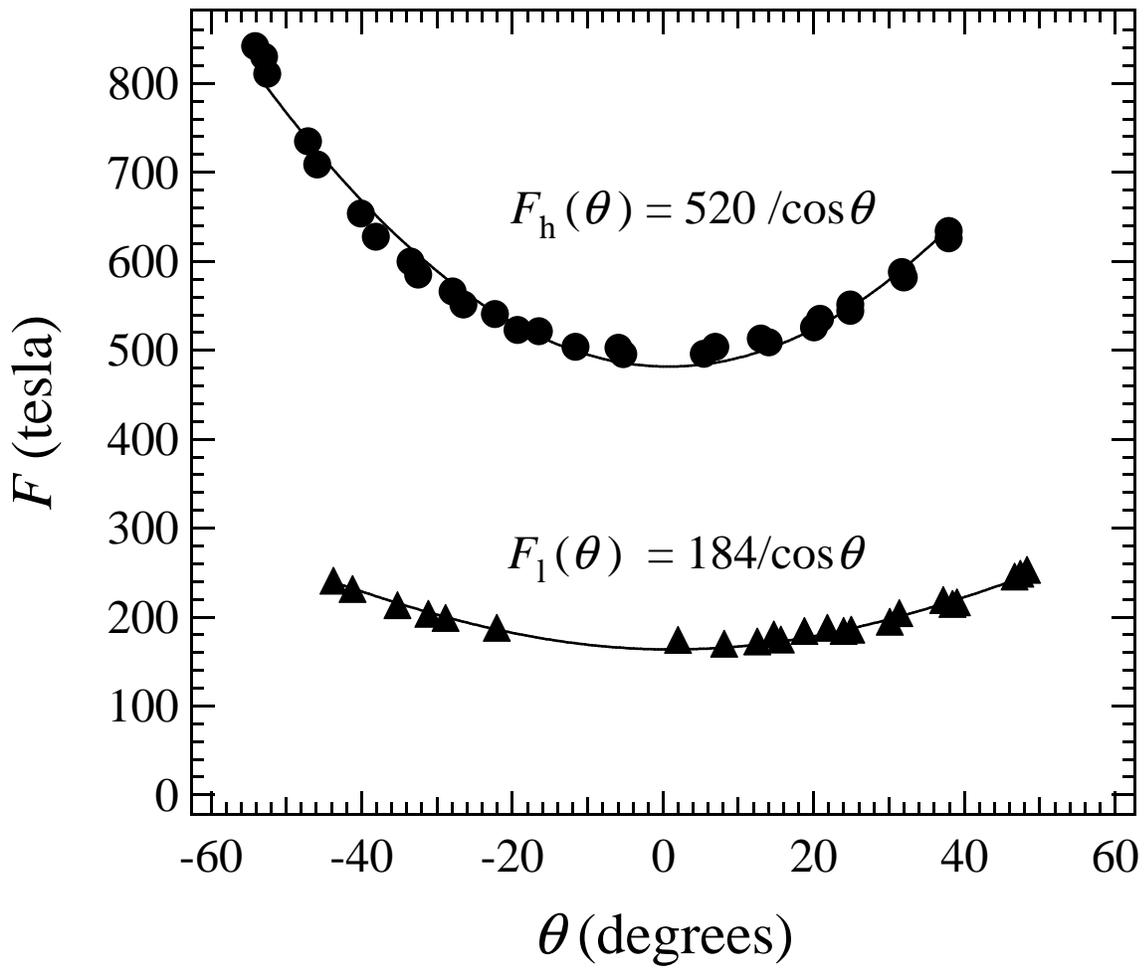
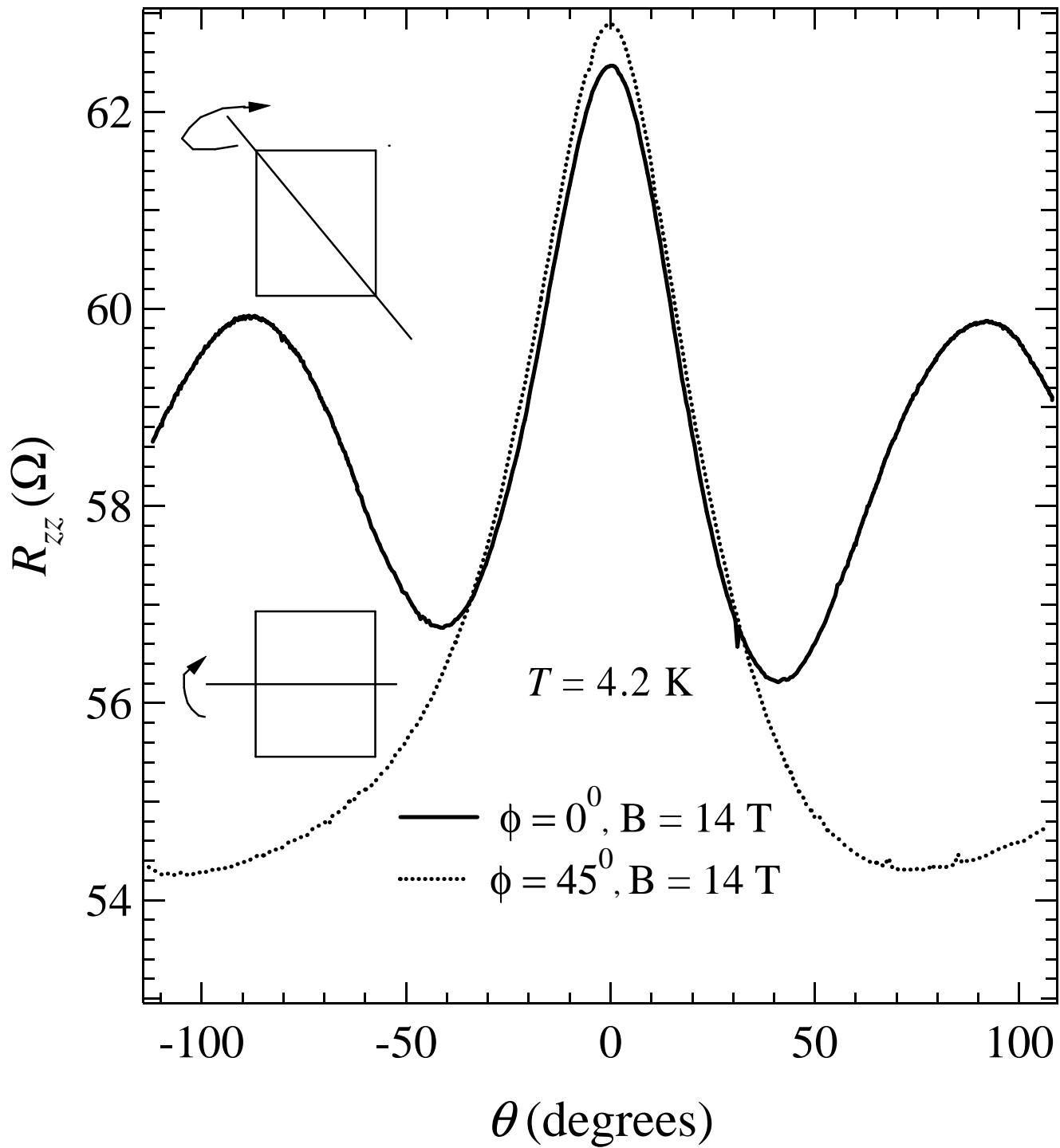
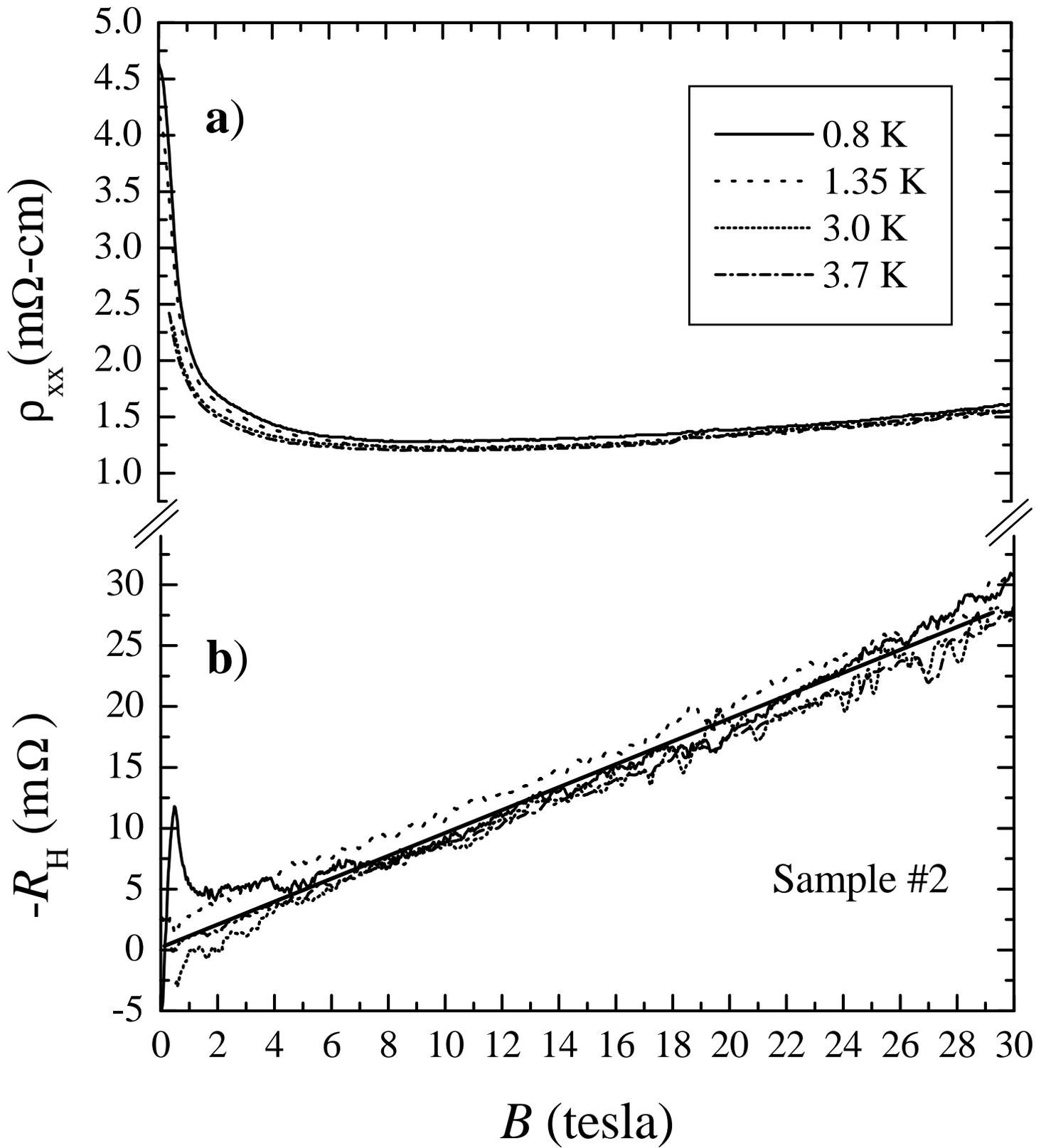
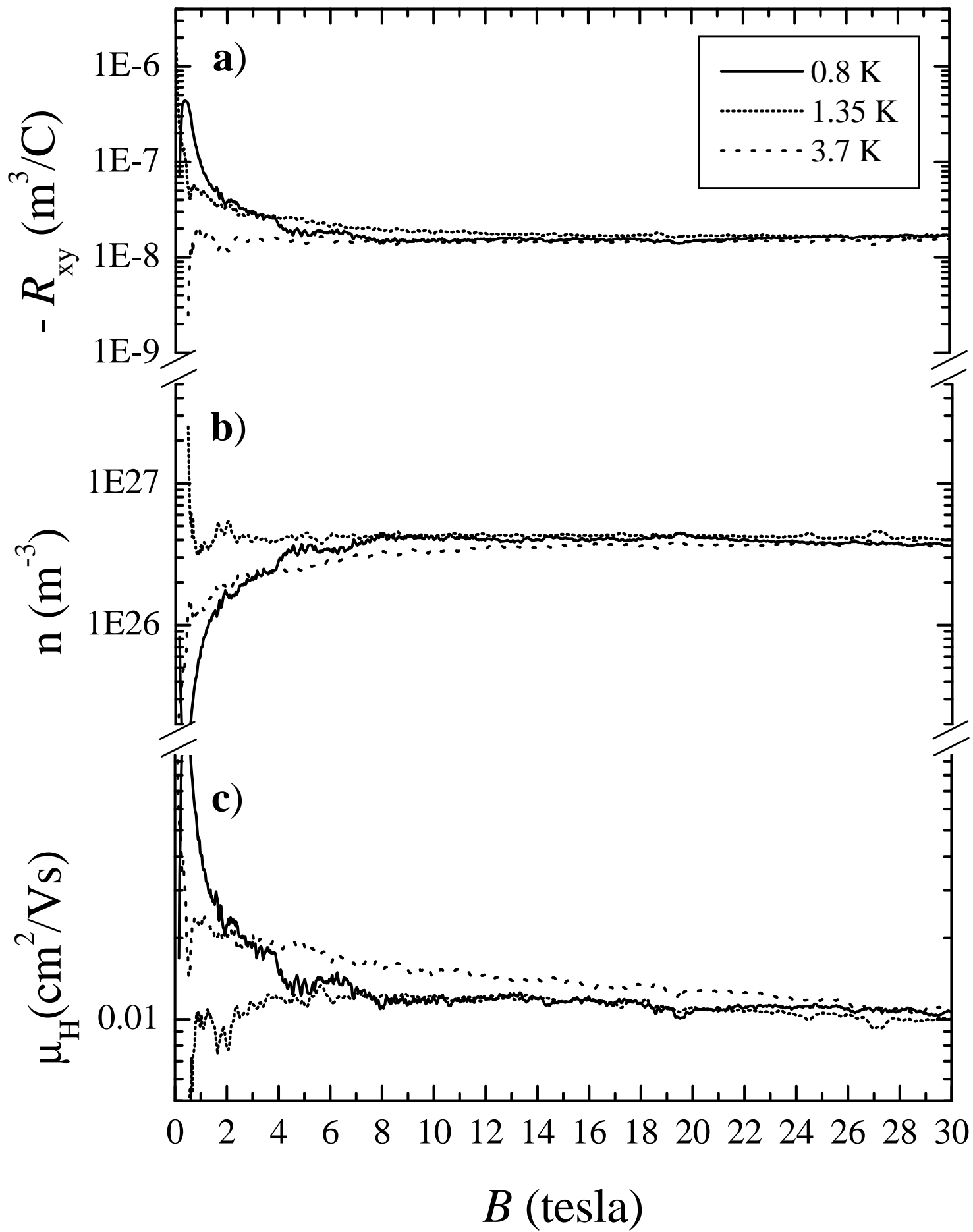
\end{document}